# Theoretical prediction of a two-dimensional intrinsic double-metal ferromagnetic semiconductor MnCoO$_4$

Tiantian Xiao, Guo Wang* and Yi Liao*

*Department of Chemistry, Capital Normal University, Beijing 100048, China.*

*Email:* wangguo@mail.cnu.edu.cn, yliao@cnu.edu.cn

ABSTRACT: A two-dimensional double-metal oxide MnCoO$_4$ was predicted to be an intrinsic ferromagnetic semiconductor by using density functional theory. The low cleavage energy 0.36 J·m$^{-2}$, which is similar to that of graphene, indicates that it can be easily exfoliated. The bulk structure has an antiferromagnetic ground state while the ferromagnetic configuration is the ground state against two antiferromagnetic and three ferrimagnetic configurations in the two-dimensional structure. The spin flip gaps for valence and conduction bands are 0.41 and 0.10 eV calculated with the HSE06 density functional, which are much larger than the thermal energy at room temperature. The Curie temperature obtained from the Monte Carlo simulation is 40 K. Under 9% tensile strain, the spin flip gaps increase largely so that the spin flip can be suppressed. The direct antiferromagnetic coupling between the Mn and Co atoms reduces largely while the indirect ferromagnetic couplings between two Mn or two Co atoms mediated by the O atoms do not decrease much in the stretched structure. The Curie temperature increases to 230 K, higher than the dry ice temperature. Moreover, phonon dispersion indicates that the MnCoO$_4$ is also stable under the tensile stain. Therefore, two-dimensional MnCoO$_4$ could be a good candidate for low-dimensional spintronics.

## 1. Introduction

According to Moore's law, the size of an electronic device becomes progressively smaller with the fast developing of computer industry. The power consumption and thermal control of electronic devices become severer problems that limit its development. Spintronics, in which spin plays the role of electric charge in traditional electronics, is a promising topic that can help to reduce the power consumption [1]. Unlike traditional devices, magnetic materials are required in spintronic devices. The source and drain electrodes should be magnetic metals or preferential half-metals. There are known materials, such as Heusler alloys [2], which are half-metallic. Since the exfoliation of graphene [3], two-dimensional materials are also good candidates for electronic devices, not only because of their appealing electronic properties but also the possibility of reducing the size of electronic devices. Besides many predicted two-dimensional structures, there are two-dimensional materials that can be likely exfoliated from known layered solids [4]. By searching the materials in the database, two-dimensional FeX$_2$ (X = Cl [5], Br and I) were predicted to be stable half-metals [6]. Thus FeX$_2$ is a candidate for electrodes in spintronic devices.

For the central scattering region of a spintronic device, a magnetic semiconductor is required. Unfortunately, magnetism in semiconductors at high temperature, especially at room temperature, is a big question facing science [7]. Although a magnetic semiconductor $CrBr_3$ has been reported in 1960 [8], the Curie temperature is only 37 K. Much effort has been made in the past decades to increase the Curie temperature of magnetic semiconductors. Dilute ferromagnetic semiconductors are promising candidates [9,10] and there are two-dimensional doped materials [11-16]. Here, we focus on intrinsic magnetic semiconductors. Bulk $La_2NiMnO_6$ is a near room temperature ferromagnetic semiconductor [17]. For two-dimensional structures, monolayers, such as $CrXTe_3$ (X = Si and Ge) [18], $CrX_3$ (X = F [19], Cl, Br and I [20]), $CrPS_4$ [21], $VX_2$ (X = S, Se, Te) [22], as well as bilayer and trilayer of δ-FeOOH [23] were predicted to be intrinsic magnetic semiconductors. These materials that contain different metal atoms provide various possibilities for us and can be applied to different situations. After searching the materials in the database [4], we are interested in a two-dimensional double-metal oxide $MnCoO_4$. The bulk $MnCoO_4$ was reported to be used as semiconductor films for gas sensors [24]. The magnetic coupling between different types of metal atoms should be more complex in such a structure. Furthermore, it is an oxide that may be more stable against oxygen in air. In the present work, two-dimensional and the bulk form of $MnCoO_4$ were investigated by using density functional theory. The electronic and magnetic properties of the two-dimensional $MnCoO_4$ were discussed in detail.

## 2. Models and computational details

The side view and top view of the two-dimensional $MnCoO_4$ are shown in Fig. 1(a)

and (b). It can be seen that the Mn and Co atoms are hexa-coordinated. The metal atoms are sandwiched between two O layers. As indicated by a rectangle in Fig. 1(b), the MnCoO$_4$ has a rectangular unit cell. A unit cell contains a Mn, a Co and four O atoms. The Mn and Co atoms extend along the X direction, while the two types of metal atoms alternate along the Y direction. For this double-metal oxide, complex magnetic couplings were investigated. Besides the ferromagnetic (FM) configuration (the spins on all the Mn and Co atom are parallel), antiferromagnetic (AFM) and ferrimagnetic (FIM) configurations were also considered. In the AFM1 configuration as shown in Fig. 1(c), all the metal atoms couples ferromagnetically along the extended line (the X direction), while both FM and AFM coupling exist along the Y direction. Oppositely, the metal atoms couple antiferromagnetically along the X direction in the AFM2 configuration shown in Fig. 1(d). The magnetic moment on a Mn and a Co atom can be different. In this case, the total magnetic moment is not zero when the spin on a Mn and a Co atom are antiparallel. Thus a FIM1 configuration (not shown in Fig. 1) can be defined when the spins on all the Mn atoms are antiparallel to that on the Co atoms. Another two configurations FIM2 and FIM3 are defined in Fig. 1(e) and (f). The Mn atoms couples antiferromagnetically while the Co atoms couples ferromagnetically along the X direction in the FIM2 configuration, whereas it is just opposite in the situation for the FIM3 configuration.

The calculations were performed with the VASP program [25]. The PBE density functional [26] and the projector-augmented wave basis with an energy cutoff of 600 eV were used. In order to describe the strongly correlated localized 3d electrons of the transition-metal atoms, the GGA+U method was used. The U–J values for the 3d electrons of Mn and Co atoms were set to 4.0 and 3.3 eV [27]. Spin-polarization was considered throughout the calculations since there are 3d electrons. For the two-dimensional structures, a vacuum layer with thickness of about 15 Å was applied along the non-periodic direction to avoid the interaction between image structures. A Monkhorst−Pack k-point sampling with 45 and 25 points along the X and Y directions in the first Brillouin zone was used. The k-point density is about 0.05 Å$^{-1}$. The phonon dispersion was calculated under density functional perturbation theory with the aid of the Phonopy code [28].

## 3. Results and discussions

*3.1. Structures and electronic properties*

After geometric optimizations, the energies of the AFM1, AFM2, FIM1, FIM2, and FIM3 configurations with respect to that of the FM configuration are 9, 20, 29, 31, and 2 meV per primitive cell. The FM configuration is the ground state. In the FM configuration, the lattice parameters of the optimized unit cell are 2.88 and 4.97 Å, respectively. Thus the distances between the nearest Mn (Co) atoms is 2.88 (4.97) Å along the X (Y) direction. The bond length between the Mn (Co) and O atoms is around 1.93 (1.90) Å. The distance between nearest Mn and Co atoms is 2.87 Å. Because the metal atoms are hexa-coordinated and each O atom is bonded to three metal atoms, the formal charges of the Mn and Co atoms are both four. Thus there are three or five d electrons left in a Mn or Co atom. The calculation indicates that the total magnetic moment for a primitive cell is 4 $\mu_B$. The magnetic moments on a Mn, Co, and O atoms are 3.2, 1.2, and −0.1 $\mu_B$, respectively. Compared with the FM configuration, the AFM and FIM configurations have little change in the lattice parameters and the bond lengths. The change is no more than 0.01 Å. The energy difference should come mainly from the magnetic coupling.

The band structures of the FM configurations are shown in Fig. 2(a). Because of the non-zero total magnetic moment, the band structures for the majority and minority spins are different. Indirect band gap exists for both spins, so the two-dimensional $MnCoO_4$ is a ferromagnetic semiconductor. The band gaps for both spins are 1.56 and 1.62 eV, respectively. For a magnetic semiconductor, spin flip should be avoided as much as possible. The spin flip gap for the valence (conduction) band can be calculated as the energy difference between the valence band maximums (conduction band minimums) for both spins. Thus the spin flip gaps in Fig. 2(a) are 0.07 and 0.01 eV. These values are not large. However, the band structures are also calculated with the HSE06 density functional [29], which can accurately describe the band gaps for of solids [30]. As shown in Fig. 2(b), the band structures calculated with the HSE06 density functional also indicate that the two-dimensional $MnCoO_4$ is a magnetic semiconductor. The total magnetic moment does not change. These confirm the results calculated with the PBE density functional. The spin flip gaps for the valence and conduction bands in Fig. 2(b) are 0.41 and 0.10 eV, respectively. These values are much larger than the thermal energy at room temperature (0.026 eV), so spin flip can be largely suppressed.

Projected density of states near the Fermi level was drawn in Fig. 3 to get insight into the bonding characteristics. The total density of states near the Fermi level is mainly contributed by the d orbitals of the Mn and Co atoms, as well as the p orbital of the O atoms. There are many peaks in Fig. 3. The peaks of the three types of the orbitals exist at the same energy window, indicating that bonds are formed between these orbitals of the atoms. There is no localized state near the Fermi level. The wide peaks indicate that the three types of orbitals are delocalized near the Fermi level. The itinerant characteristics of electrons (especially for the d electrons of the metal atoms) should be favorable to ferromagnetism [31]. Because there are four O atoms in a primitive cell, the density of states for the O atoms is larger than the others. It is noted that the density of states for the majority spin of the Mn atom is comparable to that of the Co atom, while the density of states for the minority spin of the Mn atom is quite smaller than that of the Co atom. The reason should be related to the magnetic moments. Because the magnetic moment on the Mn and Co atoms are 3.2 and 1.2 $\mu_B$, respectively, there are roughly both three electrons with majority spin while zero and two electrons with minority spin for the two types of atoms.

*3.2. Cleavage energy*

Two-dimensional materials began to be noticed since graphene was exfoliated. Low cleavage energy is essential for the success of exfoliation. For this reason, three-dimensional bulk $MnCoO_4$ [32] was also calculated. The bulk $MnCoO_4$ has a layered structure with weak interaction between O atoms in adjacent layers. The optPBE-vdW density functional [33,34], which accounts for dispersion interactions, was used to calculate the three-dimensional $MnCoO_4$ and also the two-dimensional one for comparison. Besides the above two AFM and three FIM configurations, another AFM configuration AFM3 was also considered, in which the metal atoms couples ferromagnetically inside the two-dimensional layer while antiferromagnetically along the Z direction. The energies of the AFM1, AFM2, AFM3, FIM1, FIM2, and FIM3 configurations with respect to that of the FM configuration are –38, 12, –9, 43, –20, and –5 meV per primitive cell. Thus the bulk has an AFM1 ground state and is not a ferromagnetic semiconductor. The cleavage energy calculated based on this configuration is 0.36 J·m$^{-2}$, which is the same with that of graphene (deduced from 61 meV·atom$^{-1}$) [35] and is comparable to the values 0.35-0.38 J·m$^2$ of $CrXTe_3$ [18] and 0.28-0.30 J·m$^2$ of $CrX_3$ [19]. The low cleavage energy indicates that the two-dimensional $MnCoO_4$ can be exfoliated.

*3.3. Phonon dispersion and molecular dynamics*

Phonon dispersion was calculated to investigate the dynamic stability of the two-dimensional $MnCoO_4$. A 4×3 supercell was used so that the lattice parameters are larger than 10 Å. In Fig. 4(a), there is no non-trivial imaginary frequency. The three small imaginary frequencies at the Γ point are around 4 cm$^{-1}$ and are related to three translational modes [36]. The phonon dispersion confirms the stability of the two-dimensional $MnCoO_4$. Molecular dynamics was also performed to investigate its thermal stability. A canonical ensemble and a 2×2 supercell were used. The simulation temperature is 300 K and the step size was set to 1 fs. In Fig. 5, the energy oscillates up and down many times in the 5000 steps simulation, which indicates that equilibrium has been reached. After the simulation, no essential structural deformation was observed. These imply that the two-dimensional MnCoO$_4$ is stable against heat. Although the total magnetic moment per primitive cell is also 4 μ$_B$ during the simulation, it does not mean that the FM configuration is sustainable at room temperature. The converged configuration depends on the initial configuration. Monte Carlo simulation [37,38] should be more appropriate.

*3.4. Curie temperature*

Before the Monte Carlo simulation, exchange parameters need to be calculated from the Hamiltonian based on the Heisenberg model:

$$H = -\sum_{i,j} J_{Mn-Mn} S_i \cdot S_j - \sum_{k,l} J_{Co-Co} S_k \cdot S_l - \sum_{m,n} J_{Mn-Co} S_m \cdot S_n$$

in which $J$ is an nearest-neighbor exchange parameter and $S$ is the spin on an atom. The $S$ for a Mn or Co atom is set to 3/2 or 1/2. Following this model, the energy functional per primitive cell can be written as follows:

$$E_{FM} = -\frac{9}{4} J_{Mn-Mn} - \frac{1}{4} J_{Co-Co} - 3 J_{Mn-Co} + E_0$$

$$E_{AFM1} = -\frac{9}{4} J_{Mn-Mn} - \frac{1}{4} J_{Co-Co} + E_0$$

$$E_{AFM2} = \frac{9}{4} J_{Mn-Mn} + \frac{1}{4} J_{Co-Co} + E_0$$

$$E_{FIM1} = -\frac{9}{4} J_{Mn-Mn} - \frac{1}{4} J_{Co-Co} + 3 J_{Mn-Co} + E_0$$

$$E_{FIM2} = \frac{9}{4} J_{Mn-Mn} - \frac{1}{4} J_{Co-Co} - \frac{3}{2} J_{Mn-Co} + E_0$$

$$E_{FIM3} = -\frac{9}{4} J_{Mn-Mn} + \frac{1}{4} J_{Co-Co} - \frac{3}{2} J_{Mn-Co} + E_0$$

Thus, the exchange parameters can be obtained by the following equations:

$$J_{Mn-Mn} = (-E_{AFM1} + E_{AFM2} + E_{FIM2} - E_{FIM3})/9$$

$$J_{Co-Co} = -E_{AFM1} + E_{AFM2} - E_{FIM2} + E_{FIM3}$$

$$J_{\text{Mn-Co}} = ( E_{\text{FIM1}} - E_{\text{FM}} ) / 6$$

In these equations, the energies of the AFM1, AFM2, FIM1, FIM2, and FIM3 configuration should be those under the geometry of the FM configuration. These values are 10, 23, 31, 32, and 4 meV, respectively. They have little difference with those under the optimized geometries, because changes of the geometries are quite small which is described above. The exchange parameters $J_{\text{Mn-Mn}}$, $J_{\text{Co-Co}}$, and $J_{\text{Mn-Co}}$ are 4, −15, and 5 meV, respectively. This implies that weak FM couplings exist between the same type of metal atoms while a stronger AFM coupling exists between the Mn and Co atoms. In order to analysis this phenomenon, an illustration based on the crystal field theory is presented in Fig. 6. Because the metal atoms in the two-dimensional $MnCoO_4$ are hexa-coordinated, the five d orbitals are split into $t_{2g}$ and $e_g$ orbitals. The direct FM coupling can not occur between two electrons in the $t_{2g}$ orbitals of the Mn atoms, since the $t_{2g}$ orbitals are half-occupied. The FM coupling can only occur between an electron in a $t_{2g}$ orbital and another in an $e_g$ orbital with higher orbital energy. Thus the exchange parameter $J_{\text{Mn-Mn}}$ is small. It is noticed that the spin on the O atom is antiparallel to those on the Mn atoms. The spin density shown in Fig. 7 also clearly indicates this point. The FM coupling can be an indirect superexchange interaction mediated by the p orbitals of the O atoms. The situation for $J_{\text{Co-Co}}$ is similar. On the contrary, the AFM coupling between the Mn and Co atoms results in fully occupied $t_{2g}$ orbitals. Thus the absolute value of $J_{\text{Mn-Co}}$ is larger than the other two. The AFM coupling can be direct.

After calculating the exchange parameters, the Monte Carlo simulation was performed with $10^8$ loops at a certain temperature. An 80×80 supercell was used and the number of metal atoms is in the order of $10^4$. In each loop, the spin on a metal atom flip randomly. The averaged total magnetic moment is shown in Fig. 8. The Curies temperature is estimated to be 40 K. This is rather low for practical application. Stronger FM couplings are essential for higher Curie temperature. Usually, a two-dimensional material is placed on a substrate. The mismatch of lattice parameters can apply strain to the material. The exchange interaction, especially for the direct exchange, depends significantly on the distance of atoms [18]. The two-dimensional $MnCoO_4$ with external stain is also investigated.

*3.5. Under tensile strain*

Under 1% compressive strain, the energies of the FM and FIM3 configurations are almost degenerate. The FIM3 configuration becomes the ground state under 2% compressive strain. On the contrary, FM couplings are enforced and the relative energies of the AFM and FIM configurations increase when tensile strain is applied. The $J_{Mn-Co}$ becomes positive when 7% tensile strain is applied. This means that the coupling between the Mn and Co atoms becomes FM at this time. The exchange parameters $J_{Mn-Mn}$, $J_{Co-Co}$, and $J_{Mn-Co}$ increase to 12, 11, and 21 meV, respectively, when 9% tensile strain is applied. The reason should be that the direct AFM coupling is largely suppressed when the distances of atoms increase while the indirect FM couplings mediated by O atoms do not decrease much. The FM configuration has not been obtained when 10% tensile strain is applied. The Curie temperatures are 40, 50, 70, 70, 90, 110, 170, and 190 K when 1%-8% tensile strains are applied. As shown in Fig. 8, the Curie temperature reaches 230 K when 9% tensile strain is applied. This is interesting because the temperature is higher than the dry ice temperature. This value is higher than those of the stretched $CrGeTe_3$ (108.9 K) [18] and stretched $CrI_3$ (130 K) [19], but is lower than the above room temperature for doped $CrCl_3$ [20] and $VX_2$ [22]. The next question is whether the stretched structure is stable. Fortunately, the phonon dispersion shown in Fig. 4(b) also has no non-trivial imaginary frequency. The three small imaginary frequencies at the $\Gamma$ point are around 7 $cm^{-1}$ and are also related to three translational modes. The phonon dispersion indicates that the two-dimensional $MnCoO_4$ with 9% tensile strain is stable. The strain is relaxed by expansion along the Z direction. The thickness reduces from 1.91 to 1.71 Å. The band structures shown in Fig. 2(c) also indicate that it is a magnetic semiconductor. The total magnetic moment remains 4 $\mu_B$. What is different is that the spin flip gap changes under 9% tensile strain. The spin flip gap for the valence (conduction) band calculated with the PBE density functional largely increases from 0.07 (0.01) to 0.52 (1.06) eV. Thus the spin flip is suppressed when 9% tensile strain is applied.

## 4. Conclusions

A two-dimensional double-metal oxide $MnCoO_4$ was predicted to be an intrinsic ferromagnetic semiconductor by using density functional theory. Complex magnetic couplings exist in this double-metal oxide. Besides the FM configuration, two AFM and three FIM configurations were also investigated. The two-dimensional $MnCoO_4$ has a FM ground state. The band structures indicate that it is an intrinsic ferromagnetic semiconductor. The spin flip gaps for valence and conduction bands are 0.41 and 0.10 eV calculated with the HSE06 density functional, which are much larger than the thermal energy at room temperature so that the spin flip can be suppressed. Detailed analysis indicates that the d orbitals of the Mn and Co atoms as well as the p orbital of the O atoms contribute mainly to the total density of states near the Fermi level. Indirect FM coupling mediated by O atoms exists between two Mn or two Co atoms, while direct AFM coupling exists between the Mn and Co atoms. The exchange parameter $J_{\text{Mn-Mn}}$, $J_{\text{Co-Co}}$, and $J_{\text{Mn-Co}}$ were calculated to be 4, –15, and 5 meV, respectively. The Curie temperature calculated by the Monte Carlo simulation is 40 K. The cleavage energy is 0.36 J·m$^{-2}$, which is the same with that of graphene. This implies that the two-dimensional MnCoO$_4$ should be easily exfoliated. Its stability was also confirmed by phonon dispersion and molecular dynamics. Furthermore, the Curie temperature increases to 230 K when 9% tensile strain is applied. The spin flip gaps also increase largely. Phonon dispersion indicates that the stretched MnCoO$_4$ is stable. The Curies temperature is above the dry ice temperature and the two-dimensional MnCoO$_4$ should be a good candidate for low-dimensional spintronics.

## Conflicts of interest

None.

## Acknowledgements

This work was supported by the National Natural Science Foundation of China (21203127) and the Scientific Research Base Development Program of the Beijing Municipal Commission of Education.

**References**

[1] S.A. Wolf, D.D. Awschalom, R.A. Buhrman, J.M. Daughton, S. von Molnár, M.L. Roukes, A.Y. Chtchelkanova, D.M. Treger, Spintronics: a spin-based electronics vision for the future, Science 294 (2001) 1488.

[2] R. A. de Groot, F. M. Mueller, P. G. van Engen, K. H. J. Buschow, New class of materials: half-metallic ferromagnets, Phys. Rev. Lett. 50 (1983) 2024.

[3] K. S. Novoselov, D. Jiang, F. Schedin, T. J. Booth, V. V. Khotkevich, S. V. Morozov, A. K. Geim, Two-dimensional atomic crystals, Proc. Natl. Acad. Sci. U.S.A. 102 (2005) 10451.

[4] M. Ashton, J. Paul, S. B. Sinnott, R. G. Hennig, Topology-scaling identification of layered solids and stable exfoliated 2D materials, Phys. Rev. Lett. 118 (2017) 106101.

[5] E. Torun, H. Sahin, S. K. Singh, F. M. Peeters, Stable half-metallic monolayers of $FeCl_2$, Appl. Phys. Lett. 106 (2015) 192404.

[6] M. Ashton, D. Gluhovic, S. B. Sinnott, J. Guo, D. A. Stewart, R. G. Hennig, Two-dimensional intrinsic half-metals with large spin gaps, Nano Lett. 17 (2017) 5251.

[7] D. Kennedy, C. Norman, What don't we know?, Science 309 (2005) 75.

[8] I. Tsubokawa, On the magnetic properties of a $CrBr_3$ single crystal, J. Phys. Soc.

Jpn. 15 (1960) 1664.

[9] T. Dietl, a ten-year perspective on dilute magnetic semiconductors and oxides, Nature Mater. 9 (2010) 965.

[10] T. Dietl, H. Ohno, Dilute ferromagnetic semiconductors: physics and spintronic structures, Rev. Mod. Phys. 86 (2014) 187.

[11] X. Zhao, T. Wang, G. Wang, X. Dai, C. Xia, L. Yang, Electronic and magnetic properties of 1T-HfS2 by doping transition-metal atoms, Appl. Surf. Sci. 383 (2016) 151.

[12] M. Sun, S. Wang, Y. Du, J. Yu, W. Tang, Transition metal doped arsenene: A first-principles study, Appl. Surf. Sci. 389 (2016) 594.

[13] G. Li, Y. Zhao, S. Zeng, J. Ni, The realization of half-metal and spin-semiconductor for metal adatoms on arsenene, Appl. Surf. Sci. 390 (2016) 60.

[14] Q. Zhao, Z. Xiong, L. Luo, Z. Sun, Z. Qin, L Chen, N. Wu, Design of a new two-dimensional diluted magnetic semiconductor: Mn-doped GaN monolayer, Appl. Surf. Sci. 396 (2017) 480.

[15] M. Rafique, Y. Shuai, H.-P. Tan, M. Hassan, Structural, electronic and magnetic properties of 3d metal trioxide clusters-doped monolayer graphene: a first-principles study, Appl. Surf. Sci. 399 (2017) 20.

[16] N. Song, Y. Wang, W. Yu, L. Zhang, Y. Yang, Y. Jia, Electronic, magnetic properties of transition metal doped $Tl_2S$: first-principles study, Appl. Surf. Sci. 425 (2017) 393.

[17] N. S. Rogado, J. Li, A. W. Sleight, M. A. Subramanian, Magnetocapacitance and

magnetoresistance near room temperature in a ferromagnetic semiconductor: $La_2NiMnO_6$, Adv. Mater. 17 (2005) 2225.

[18] X. Lia, J. Yang, $CrXTe_3$ (X = Si, Ge) nanosheets: two dimensional intrinsic ferromagnetic semiconductors, J. Mater. Chem. C 2 (2014) 7071.

[19] W.-B. Zhang, Q. Qu, P. Zhu, C.-H. Lam, Robust intrinsic ferromagnetism and half semiconductivity in stable two-dimensional single-layer chromium trihalides, J. Mater. Chem. C 3 (2015) 12457.

[20] J. Liu, Qiang Sun, Y. Kawazoe, P. Jena, Exfoliating biocompatible ferromagnetic Cr-trihalide monolayers, Phys. Chem. Chem. Phys. 18 (2016) 8777.

[21] H. L. Zhuang, J. Zhou, Density functional theory study of bulk and single-layer magnetic semiconductor $CrPS_4$, Phys. Rev. B: Condens. Matter Mater. 94 (2016) 195307.

[22] H.-R. Fuh, C.-R. Chang, Y.-K. Wang, R. F. L. Evans, R. W. Chantrell, H.-T. Jeng, Newtype single-layer magnetic semiconductor in transition-metal dichalcogenides $VX_2$ (X = S, Se and Te), Sci. Rep. 6 (2016) 32625.

[23] I. Khan, A. Hashmi, M. U. Farooq, J. Hong, Two-dimensional magnetic semiconductor in feroxyhyte, ACS Appl. Mater. Interfaces 9 (2017) 35368.

[24] V. V. Malyshev, A. V. Pislyakov, Dynamic properties and sensitivity of semiconductor metal-oxide thick-film sensors to various gases in air gaseous medium, Sens. Actuators B 96 (2003) 413.

[25] G. Kresse, J. Furthmüller, Efficient iterative schemes for ab initio total-energy calculations using a plane-wave basis set, Phys. Rev. B: Condens. Matter Mater. Phys.

54 (1996) 11169.

[26] J. P. Perdew, K. Burke, M. Ernzerhof, Generalized gradient approximation made simple, Phys. Rev. Lett. 77 (1996) 3865.

[27] L. Wang, T. Maxisch, G. Ceder, Oxidation energies of transition metal oxideswithin the GGA + U framework, Phys. Rev. B: Condens. Matter Mater. Phys. 73 (2006) 195107.

[28] A. Togo, F. Oba, I. Tanaka, First-principles calculations of the ferroelastic transition between rutile-type and $CaCl_2$-type $SiO_2$ at high pressures, Phys. Rev. B: Condens. Matter Mater. Phys. 78 (2008) 134106.

[29] A. V. Krukau, O. A. Vydrov, A. F. Izmaylov, G. E. Scuseria, Influence of the exchange screening parameter on the performance of screened hybrid functionals, J. Chem. Phys. 125 (2006) 224106.

[30] T. M. Henderson, J. Paier, G. E. Scuseria, Accurate treatment of solids with theHSE screened hybrid, Phys. Status Solidi B 248 (2011) 767.

[31] H. Capellmann, Theory of itinerant ferromagnetism in the 3-d transition metals, Z. Phys. B: Condens. Matter Quanta 34 (1979) 29.

[32] K. Persson, 2014, doi:10.17188/1296391.

[33] J. Klimeš, D. R. Bowler, A. Michaelides, Chemical accuracy for the van der Waals density functional, J. Phys. Condens. Matter 22 (2010) 022201.

[34] J. Klimeš, D. R. Bowler, A. Michaelides, Van der Waals density functionals applied to solids, Phys. Rev. B: Condens. Matter Mater. Phys. 83 (2011) 195131.

[35] R. Zacharia, H. Ulbricht, T. Hertel, Interlayer cohesive energy of graphite from

thermal desorption of polyaromatic hydrocarbons, Phys. Rev. B: Condens. Matter Mater. Phys. 69 (2004) 155406.

[36] G. Wang, Theoretical prediction of the intrinsic half-metallicity in surface-oxygen-passivated $Cr_2N$ Mxene, J. Phys. Chem. C 120 (2016) 18850.

[37] J. Rusz, L. Bergqvist, J. Kudrnovský, I. Turek, Exchange interactions and Curie temperatures in $Ni_{2-x}MnSb$ alloys: first-principles study, Phys. Rev. B: Condens. Matter Mater. Phys. 73 (2006) 214412.

[38] X. Li, X. Wu, J. Yang, Room-temperature half-metallicity in La(Mn,Zn)AsO Alloy via element substitutions, J. Am. Chem. Soc. 136 (2014) 5664.

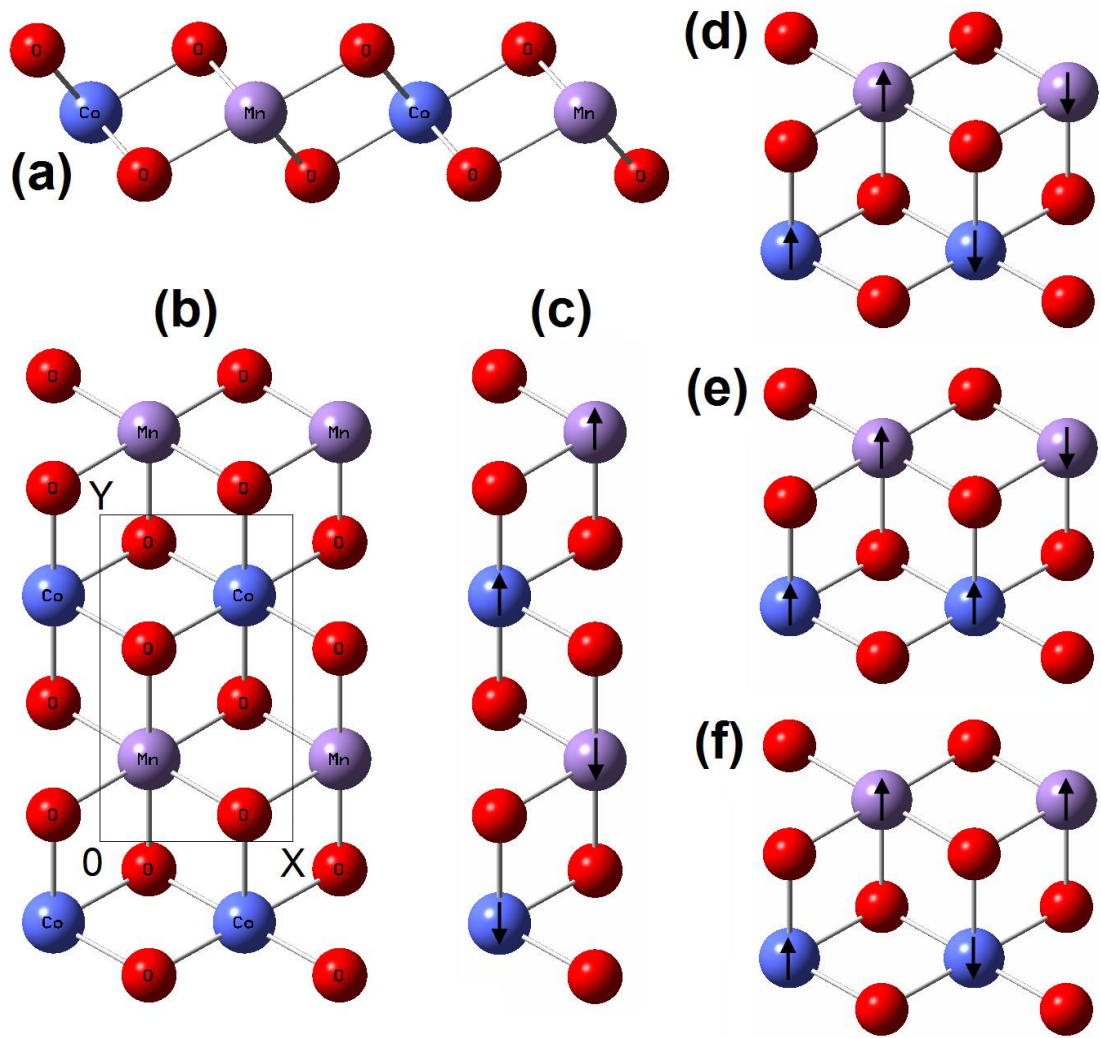

**Fig. 1** (a) Sideview and (b) top view of two-domensional MnCoO$_4$, the unit cell is denoted by a rectangle, (c) 1×2 supercell for AFM1 configuration, 2×1 supercell for (d) AFM2, (e) FIM2 and (f) FIM3 configurations. Those for FM and FIM1 configurations are not shown.

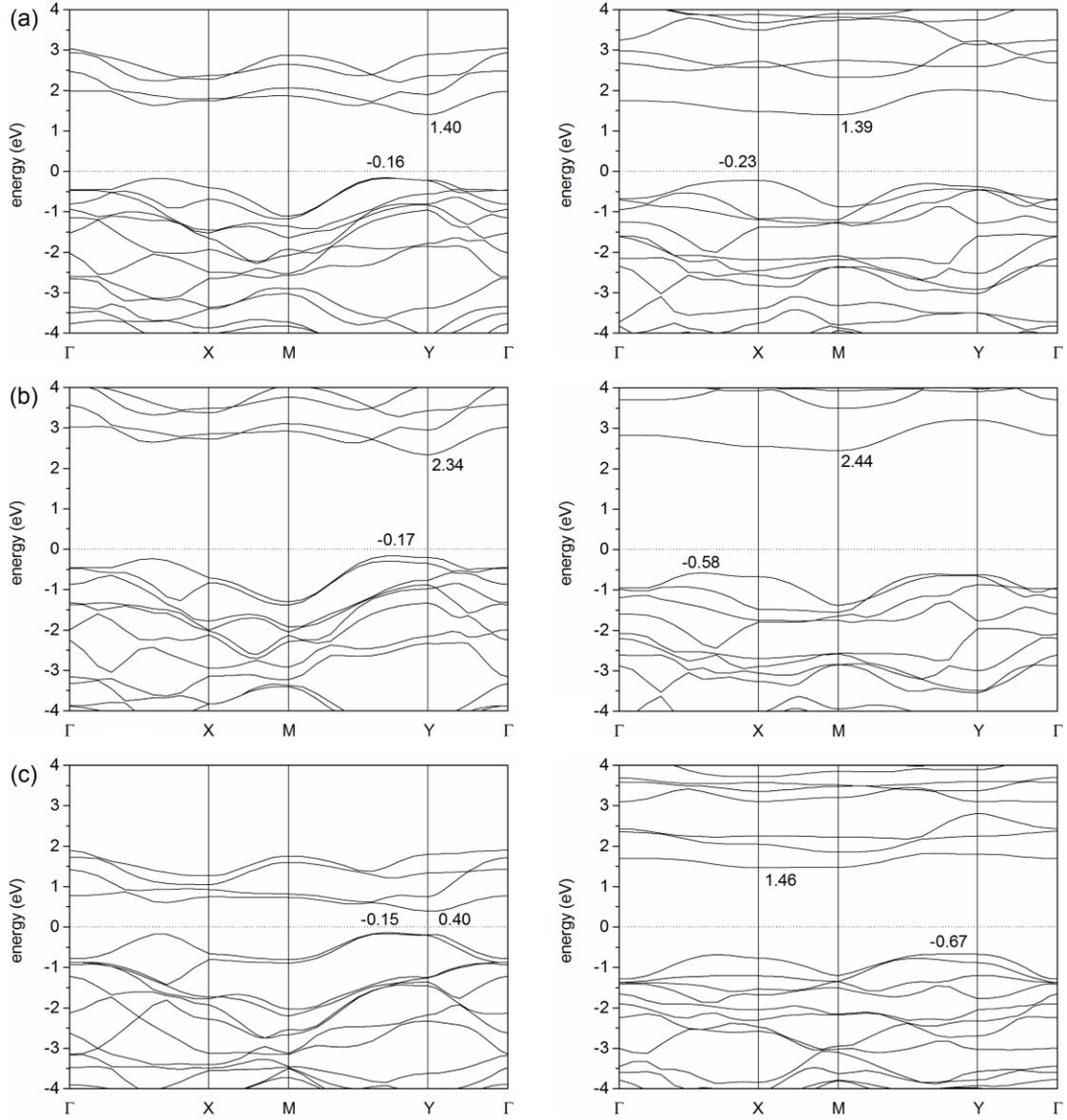

**Fig. 2** Band structures of FM configuration calculated with (a) the PBE and (b) the HSE06 density functionals, (c) band structures of FM configuration under 9% tensile strain calculated with the PBE density functional. The left is for the majority spin and the right is for the minority spin. The Fermi level is set to zero.

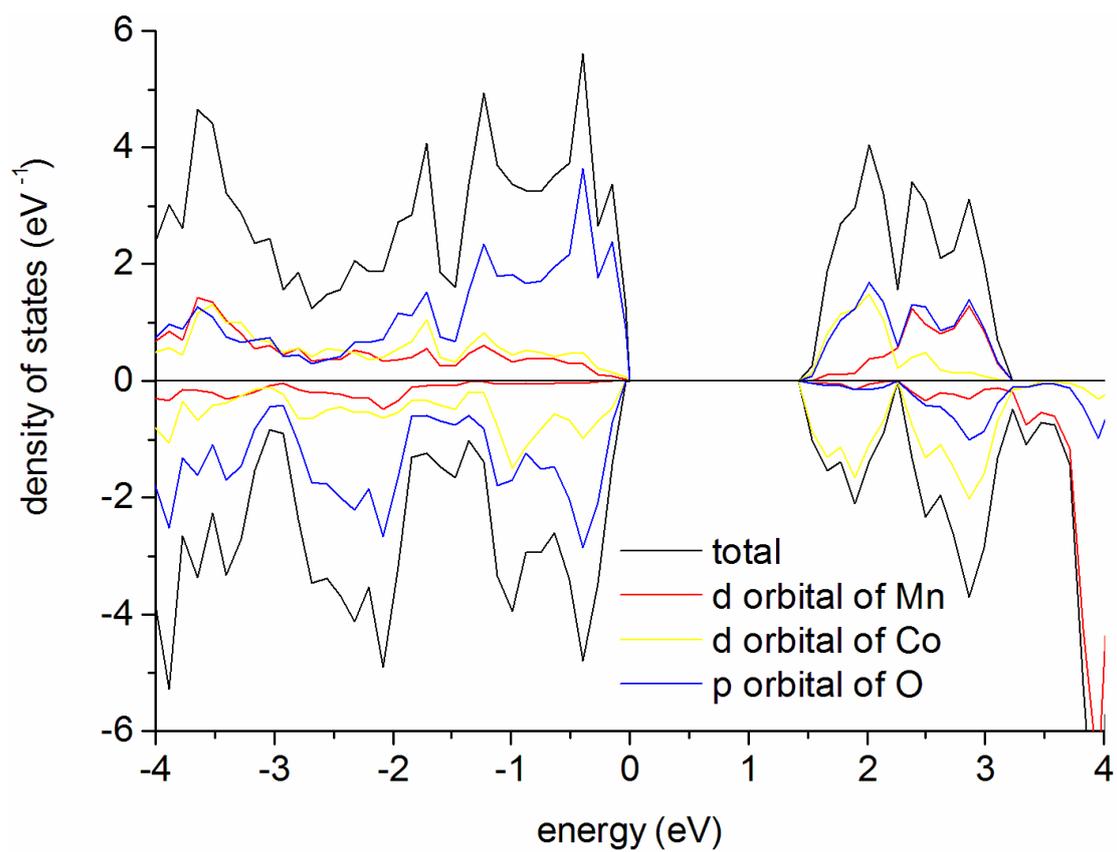

**Fig. 3** Total and projected density of states for the FM configuration of two-dimensional $MnCoO_4$.

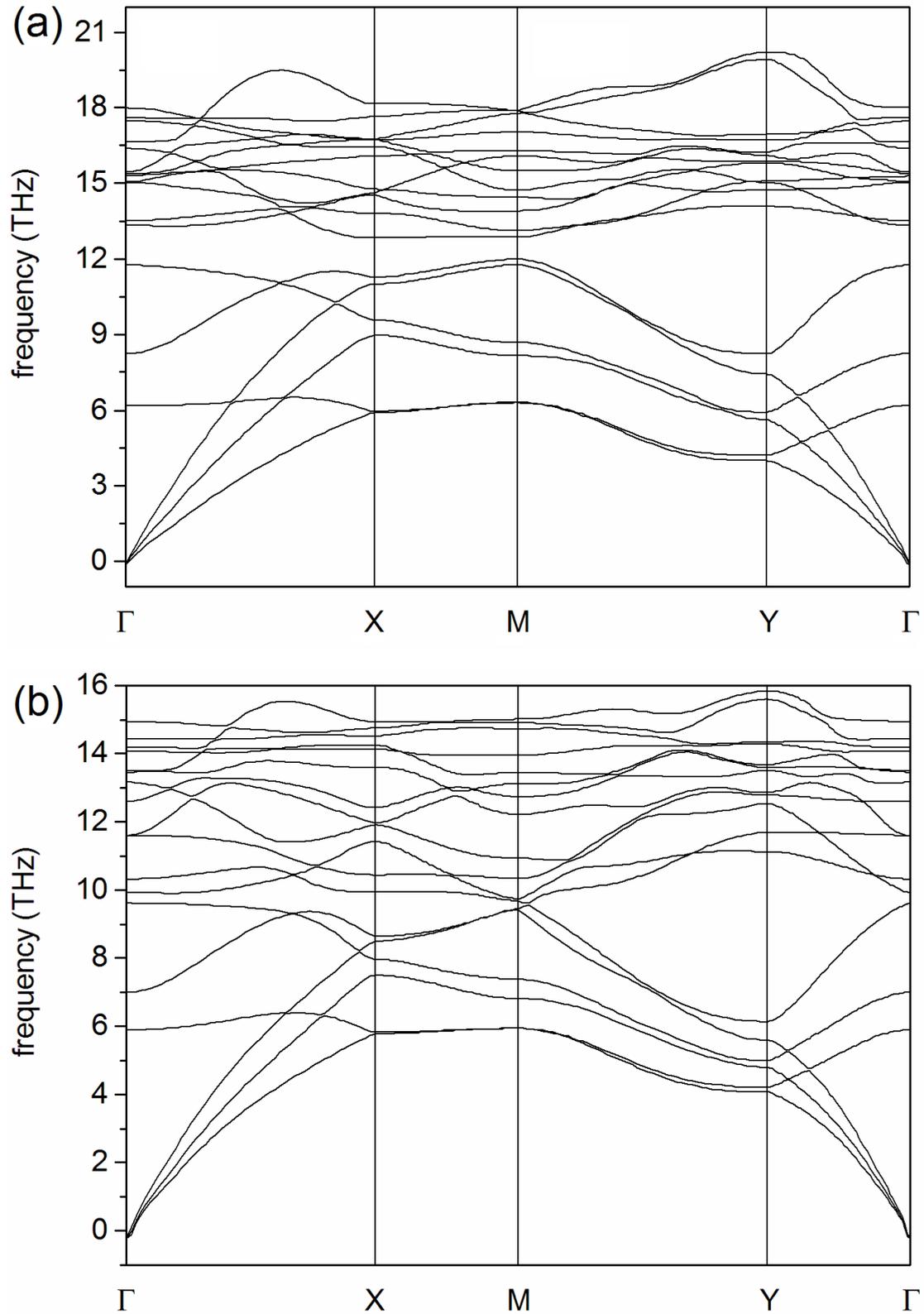

**Fig. 4** Phonon dispersions of (a) two-dimensional MnCoO$_4$ and (b) that under 9% tensile strain.

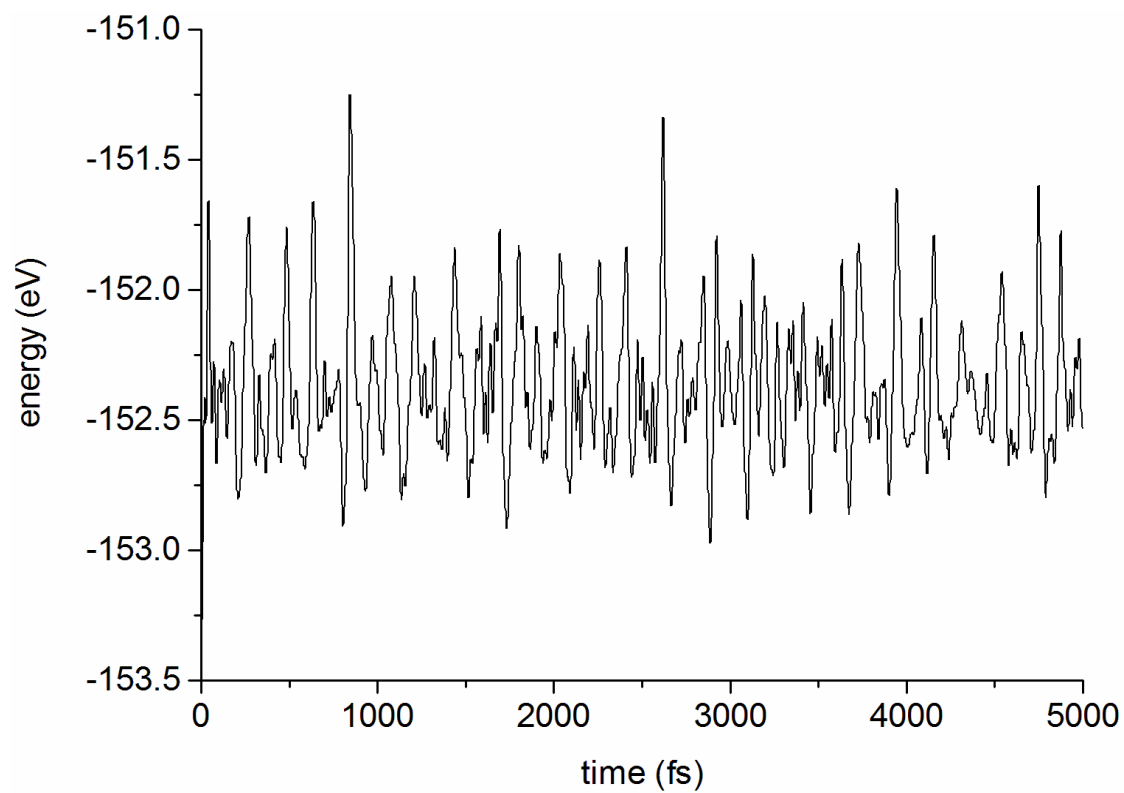

**Fig. 5** Molecular dynamics simulation of two-dimensional MnCoO$_4$.

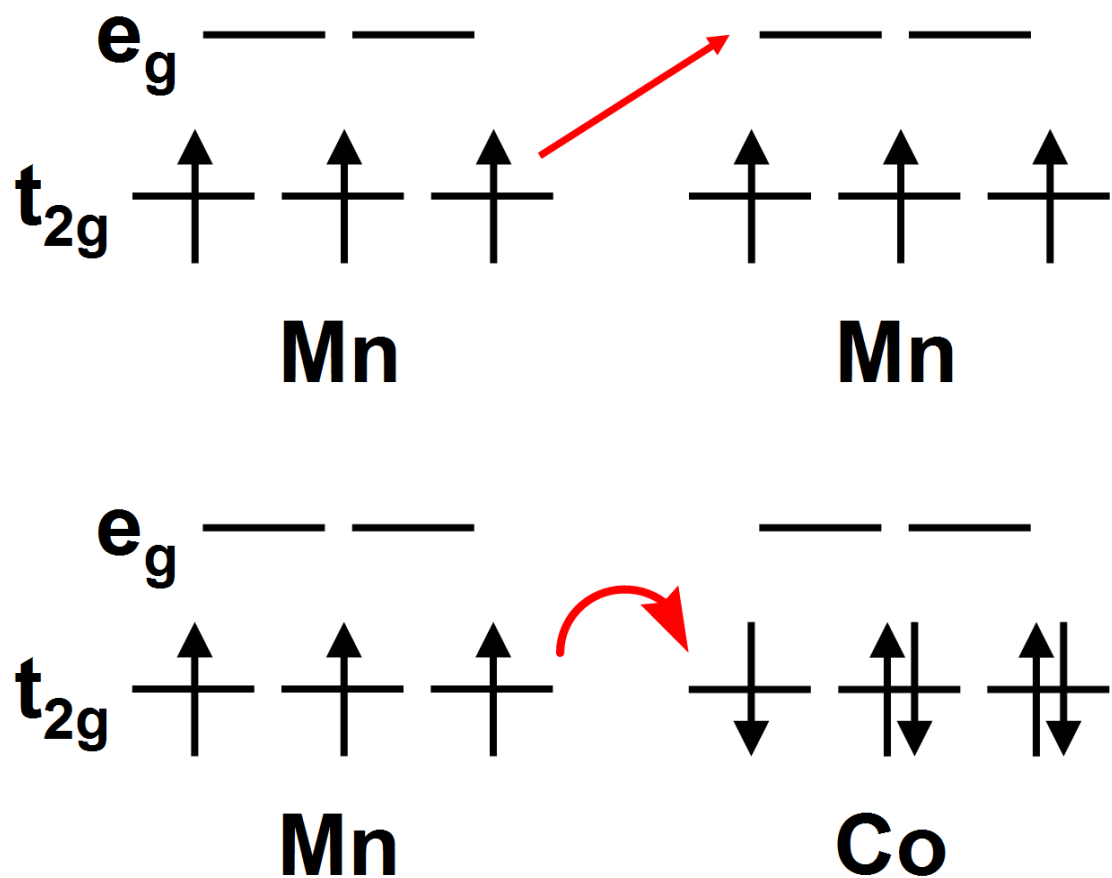

**Fig. 6** Illustration of FM coupling between two Mn atoms (top) and AFM coupling between Mn and Co atoms (bottom) in two-dimensional MnCoO$_4$.

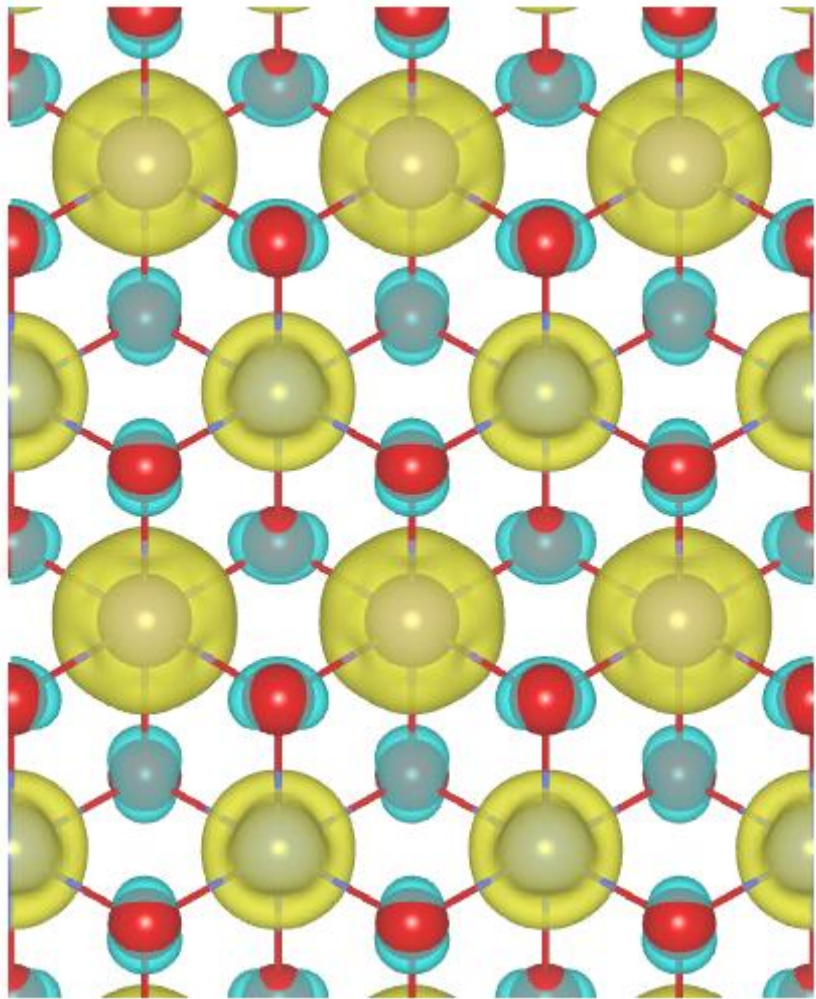

**Fig. 7** Spin density of two-dimensional MnCoO$_4$.

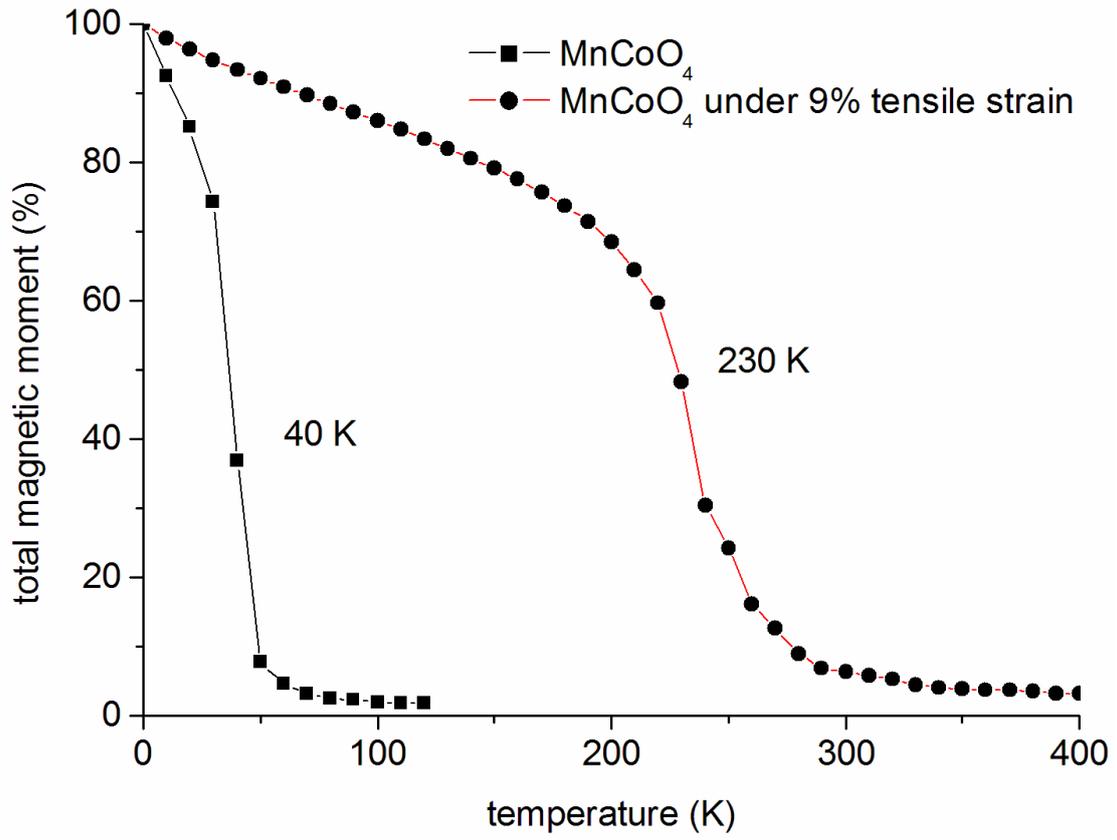

**Fig. 8** Curie temperatures of two-dimensional MnCoO$_4$ and that under 9% tensile strain obtained from Monte Carlo simulations.